\documentclass[twocolumn]{aastex63}

\usepackage{ulem}

\received{April 1, 2020}
\shorttitle{The Really Habitable Zone}
\shortauthors{Pedbost et al.}
\graphicspath{{./}{figures/}}
\begin{document}

\title{Defining the Really Habitable Zone}

\author{Marven F. Pedbost}
\affiliation{Earth}

\author{Trillean Pomalgu}
\affiliation{Earth}

%\collaboration{1}{(AAS Journals Data Scientists collaboration)}

\author{Chris Lintott}
\affiliation{Department of Physics, University of Oxford, Keble Road, OX1 3RH}
\affiliation{Lamb and Flag, 12 St Giles', Oxford, OX1 3JS}
%\affiliation{AAS Journals Associate Editor-in-Chief}
%\nocollaboration{1}

\author{Nora Eisner}
\affiliation{Department of Physics, University of Oxford, Keble Road, OX1 3RH}
%\altaffiliation{AASTeX v6+ programmer}
%\affiliation{TeXnology Inc.}

\author{Belinda Nicholson}
\affiliation{Department of Physics, University of Oxford, Keble Road, OX1 3RH}

%% Note that the \and command from previous versions of AASTeX is now
%% depreciated in this version as it is no longer necessary. AASTeX 
%% automatically takes care of all commas and "and"s between authors names.

%% AASTeX 6.3 has the new \collaboration and \nocollaboration commands to
%% provide the collaboration status of a group of authors. These commands 
%% can be used either before or after the list of corresponding authors. The
%% argument for \collaboration is the collaboration identifier. Authors are
%% encouraged to surround collaboration identifiers with ()s. The 
%% \nocollaboration command takes no argument and exists to indicate that
%% the nearby authors are not part of surrounding collaborations.

%% Mark off the abstract in the ``abstract'' environment. 
\begin{abstract}
Since the discovery of the first confirmed exoplanet, observations have revealed a remarkable diversity of worlds. A wide variety of orbital and physical characteristics are detected in the exoplanet population, and much work has been devoted to deciding which of these planets may be suitable for life. Until now, though, little work has been devoted to deciding which of the potentially habitable planets might actually be \textit{worth} existing on. To this end, we present the Really Habitable Zone (RHZ), defined as the region around a star where acceptable gins and tonic are likely to be abundant. In common with much of the work in the field, we rely throughout on assumptions which are difficult if not impossible to test and present some plots which astronomers can use in their own talks, stripped of all caveats. We suggest that planets in the Really Habitable Zone be early targets for the JWST, because by the time that thing finally launches we're all going to need a drink. 
\end{abstract}

%% Keywords should appear after the \end{abstract} command. 
%% See the online documentation for the full list of available subject
%% keywords and the rules for their use.

%% From the front matter, we move on to the body of the paper.
%% Sections are demarcated by \section and \subsection, respectively.
%% Observe the use of the LaTeX \label
%% command after the \subsection to give a symbolic KEY to the
%% subsection for cross-referencing in a \ref command.
%% You can use LaTeX's \ref and \label commands to keep track of
%% cross-references to sections, equations, tables, and figures.
%% That way, if you change the order of any elements, LaTeX will
%% automatically renumber them.
%%
%% We recommend that authors also use the natbib \citep
%% and \citet commands to identify citations.  The citations are
%% tied to the reference list via symbolic KEYs. The KEY corresponds
%% to the KEY in the \bibitem in the reference list below. 

\section{Introduction} \label{sec:intro}

Either we are alone in the Universe, or we are not. Assuming the latter, one could reasonably expect extra-terrestrial life to reside on another planet, and if this planet is outside of our own Solar System it is commonly known as an exoplanet. Since the discovery of the first exoplanet in 1995, the number of grants rewarded towards investigating this question has increased significantly. The inquiry into the existence of life, however, is an extremely complex topic involving numerous convoluted considerations, making it an ideal theme for telescope and grant funding applications, but a less practical question to answer. Instead, the community has formed a hand-shake agreement to instead investigate the more loosely defined question of habitable zones. %This comment will make the @OverheardonAph bot advertise our paper on the Really Habitable Zone. 

The Habitable Zone is defined as the region around a star where life, as we know it, could survive. This generally only considers the ability of liquid water to exist on a planet at a given distance from its host star. Although simpler than questioning the existence of life, there are still numerous factors and quantities to consider and calibrate. Due to the uncertainties and degeneracies in exoplanet sciences, as well as a lack of a decent model for convection, there exist a plethora of habitable zone models that have a tendency to disagree with one another. As such, all the research in this field has tried to find ways to define and refine the habitable zone such that the budget of JWST can be justified to US congress. 

Existing treatments of the habitable zone concept tend to be inclusive. In thinking about life in the Universe, we are necessarily forced to use the conditions life on Earth can endure as a guide \citep[e.g., ][]{horneck2012astrobiology, bada2004life}, but many have argued that this is too restrictive \citep[e.g., ][]{azua2013potential, wachtershauser2000life}. Life may be able to proceed in a variety of conditions which do not exist on Earth. Equally, it is possible that, while life on Earth is able to cope with a remarkable variety of conditions, the necessary circumstances for life to get started may be very specific, leaving many planets in today's `Habitable Zone' lifeless. These arguments underline the essential difficulty in defining a true Habitable Zone with any certainty.

However, if we can't agree on what makes life possible, surely we can agree what makes life worth living. We therefore define the Really Habitable Zone (RHZ) as the region in which a good gin and tonic is possible\footnote{Recent developments have begun the process of making non-alcoholic gin-like substitutes available; we therefore can adopt this position without implications for inclusiveness}. 

This definition makes perfect sense. Astronomers have long been interested in alcohol. Shortly after the introduction of gin to the UK, Flamsteed made the first observations of Uranus but mistook it for a star; we suggest a connection between these two events. In the twentieth century, papers by \citet{Coburn} and \citet{PhenixLittell} discuss the use of alcohols to treat photographic plates, though they remain silent on its use in treating astronomers. \citet{Ball} detected methanol, though this is - in space, as on Earth - clearly undrinkable. Luckily, \citet{Zuckerman} soon found ethanol.

It is therefore clear that conditions which support the creation of decent beverages will also support astronomers, which is a working definition of civilization. The choice of gin and tonic is based on the work of Adams \citep{Adams}, who hypothesised that a drink named something similar existed in 85\% of civilizations \footnote{We are aware that Adams added that the drinks themselves are different. As observational astronomers, we're comfortable with lumping different phenomena into a single category. We leave the division of hypothetical drinks into Type \textsc{i}, Type \textsc{iia}, Type \textsc{iib} gins and tonic for future work}. 

To proceed, we define the Minimum Acceptable Gin and tonIC, or MAGIC \citep{Cook}\footnote{As is the case for many astronomical acronyms, this may look like we were drunk when we came up with it, but we were stone cold sober.}. A MAGIC must contain: gin, tonic, ice and some sort of citrus. The citrus may be controversial, or even considered old-fashioned; we are aware of what we term the Hendricks Exception, involving `cucumber', but as neither Somerville nor New College Senior Common Rooms have yet to adapt to such modern `innovation', we can safely ignore the possibility, sticking to good old ice and a slice. In this paper we therefore consider how the properties of MAGIC affect the theoretical and observational definition of the Really Habitable Zone. 

\section{Determining the Really Habitable Zone}

Gin, in essence, is alcohol which has been flavoured with a wide variety of `botanical' species. A precise definition of `botanical' is lacking, so we assume it is the equivalent of a astronomer's use of `metal' - including almost everything in the Universe apart from a few common ingredients. Everything is a metal, apart from Hydrogen and Helium, and everything is a botanical apart from water and alcohol.

There exists a broad consensus\footnote{amongst the authors} that juniper is the essential ingredient. However, juniper can grow in an extraordinary variety of conditions and thus we should expect exojuniper to exist on a wide range of planets. There is, naturally, a taxonomic problem; any alien tree which produced juniper flavoured berries would not be members of the genus \textit{Juniperus}. However, few drinkers will care so for the purpose of this paper will call exo-Juniper `Juniper'. 

(We note the possibility of `ginspermia', in which fully formed juniper bushes or their berries are transported between planets; however, we assume efficient harvesting to enable maximal gin production means that there are no spare bushes to fly between the stars.)

This tolerance for varied climatic conditions means that the necessity of juniper does not impose strong limits on the RHZ; we note that Western Juniper, for example, exists in `soil regimes [which] are mesic and frigid (limited cryic)'\citep{Miller} \footnote{We don't understand this, but since when has that stopped anyone quoting evidence that seems to back up their argument in a paper?}. 

Juniper does, however, rely on seasonal variation in climatic conditions to fruit. There is therefore a stringent requirement that planets within the Really Habitable Zone have a significant axial tilt to be Really Habitable. This explains, for example, why there is no life worth speaking to on Mercury, which has only a very small axial tilt. By analogy with a terrible party, the lack of gin may also contribute to the planet's lack of atmosphere.

In contrast to juniper-related considerations, the region around a star where the conditions are adequate for the growing of lemons or limes, fundamental ingredients required for the gin and tonic drink, is sensitive to a number of factors. 
These necessary citrus fruits thrive in temperatures ranging from 21 to 38$^\circ$ C (botanist, priv. comm.) and require a steady supply of $\mathrm{H_2O}$, hereafter ‘water’. The planet must, therefore, carefully balance its distance from the host star with its atmospheric composition in order to meet the criterion required for these fruit to survive. While the availability of water is, of course, vital to the survival of these plants, access to dry land is also a necessity, both for the picking and slicing of the fruit, a vital step in the preparation of the gin and tonic. 

The calculation of the Really Habitable Zone is based on previous efforts used to identify the standard, or `boring', habitable zone (BHZ). \citet{Kopparapu2013, Kopparapu2014}, for example, calculate various BHZs for different planetary conditions (Recent Venus, Runaway Greenhouse, Moist Greenhouse, Maximum Greenhouse, and Early Mars) using the equation:

\begin{equation}
S_{eff} = S_{eff \bigodot} + aT_{\star} + bT_{\star}^{2} + cT_{\star}^{3} + dT_{\star}^{4}
\end{equation}

where $S_{eff}$ is the effective stellar flux and $T_{\star} = T_{eff} - 5780 K$ and the coefficients related to the model used to determine the habitable zone. The coefficients encapsulate underlying assumptions of the atmospheric modeling and the weather on the planet, as well as of the properties of the surface of the planet and other unknown planet properties that are beyond the scope of this, and indeed most, papers. We found that there is a lack of general consensus in the literature as to what these parameters should be, and an overall disagreement in the models that should be applied \citep[e.g., ][]{HBZ1, HBZ2, HBZ3, HBZ4, HBZ5, HBZ6, HBZ7, HBZ8, HBZ9, HBZ10, HBZ11, HBZ12, HBZ13, HBZ14, HBZ15}. In choosing our model parameters for the Really Habitable Zone, we therefore followed what appears to be standard practice, and made-up new coefficients entirely as shown in Table 1 according to our whims.

\begin{table*}[ht!]
\label{tab:RHBcoeffs}
\centering
\begin{tabular}{ccccccc}
\hline
          & Seff  & a        & b        & c          & d          &  \\
\hline
Inner RHZ & 1.666 & $2.136 \times 10^{-4}$ & $2.536 \times 10^{-8}$ & $-1.336 \times 10^{-11}$ & $-3.096 \times 10^{-15}$ &  \\
Outer RHZ & 0.838 & $1.886 \times 10^{-4}$ & $2.006 \times 10^{-8}$ & $-1.999 \times 10^{-11}$  & $-2.011 \times 10^{-15}$ & 
\end{tabular}
\caption{Note: all coefficients assume a pressure of 1 bar, as one simply cannot be in more than one bar at any given time. Once you find a good bar, you should stay in it.}
\end{table*}

The inner RHZ coefficients are loosely based on the `Recent Venus' coefficients derived by \citet{Kopparapu2013} due to the undoubted need for a drink on such a high pressured planet. The outer RHZ coefficients, on the other hand, were carefully derived following the consumption of multiple gins and tonic\footnote{Yes, gins and tonic, not gin and tonics. You want multiple gins, not more tonic.} (Dionysus priv. comm.). Coincidentally, these outer edge coefficients also appear to align with the founding years of some of the more well known Gin distilleries from around the world: Tanqueray London Dry Gin (1838), Bombay Sapphire London (1886), Aviation American Gin (2006), Hendrick's Gin (1999) and Botanist Islay Dry Gin (2011). 

\begin{figure}
\label{fig:RHZ_limes}
\centering
        \includegraphics[totalheight=8cm]{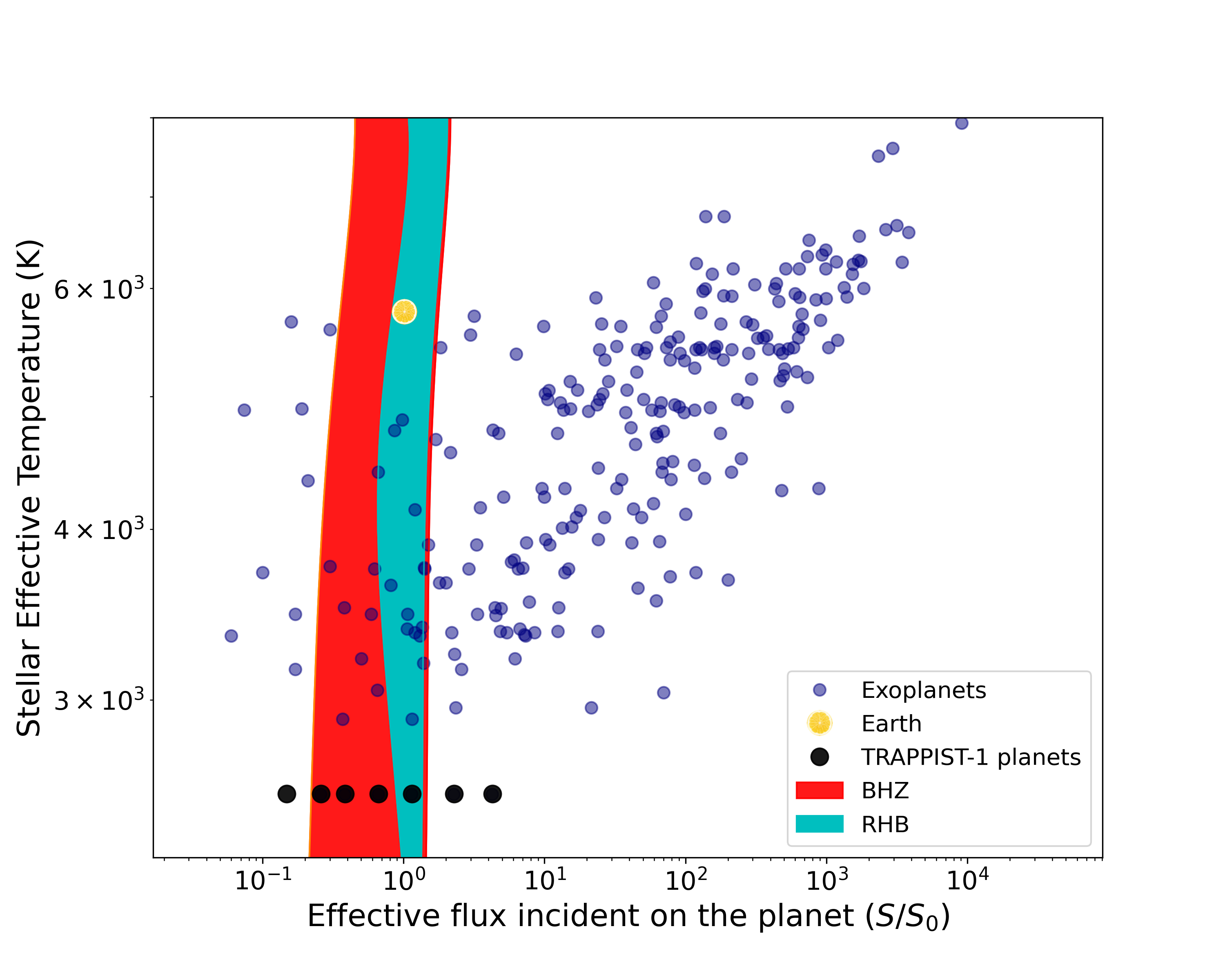}
    \caption{As this figure shows, the Earth (lemon) is worryingly close to the outermost edge of the Really Habitable Zone (teal blue region). One might think that this is cause for panic, however, the authors have extensively tested and verified the existence of gins and tonic on Earth. Our vigilence in this matter is (nearly) unceasing.}
\end{figure}

Figure 1 shows the exoplanets with known effective incident fluxes listed in the NASA Exoplanet Archive \citep[][; blue dots]{Akeson2013}, one of many possible boring habitable zones (red), and the really habitable zone (blue). As can be seen, most known exoplanets do not lie within the Really Habitable Zone, and thus can be excluded as possible homes for life. Indeed, it's hard to understand now why so much effort was put into discovering them, given their evident unsuitability. It is our hope that this publication of the RHZ will enable astronomers to stop wasting their time and concentrate on planets worth talking about. 

A much talked about exoplanet system is that of TRAPPIST-1, which consists of seven `Earth-like' planets orbiting around an ultracool dwarf star \citep{luger2017}. You might be inclined to believe that multiple of the TRAPPIST planets warrant further investigation, either because of their presence within the BHZ, or due to your associations of these planets with strong Belgian beers. However, as indicated in Figure 1, you can only find the necessary ingredients for gin and tonics on one of them, so we recommend astronomers ignore the rest forthwith. 

We had planned to carry out a statistical test to determine whether the populations of planets within and without the Really Habitable Zone were really different. However, the undergraduate student we'd expected to apply a Bayesian Machine Learning model has stopped answering our emails, so we merely suggest you follow astronomical tradition and look at the graph for a few seconds before drawing sweeping conclusions from it.

\section{Observational Prospects: Detecting gins and tonic on exoplanets}

As with many exoplanet papers, this one includes a section on wildly infeasible future observations which, though difficult, would undoubtedly have enormous scientific impact. We cannot be sure that planets within the RHZ capable of hosting gins and tonic really will have suitable beverages, and thus describe a range of possible experiments in the hope that those who carry them out in the distant future will cite this paper, burnishing our reputation for foresight and clear thinking. We will consider in turn the relevant components. 

\subsection{Gin}
We adopt the methodology described in \citet{Doughty} (D20) to consider the detection of juniper bushes, which we identify as upright-ish photosynthesising multicellular lifeforms, or UiPML. D20 rolled the dice in testing an approach to detecting the shadows of UiPML\footnote{Inventing an acronym instead of using the word 'tree' is really quite something, even for astronomers.} in footage obtained with Unmanned Aerial Vehicles in Arizona. This is exactly the same as detecting juniper on a planet several hundred light-years away, so we consider this a solved problem and will apply for telescope time immediately\footnote{The overworked Telescope Allocation Committee are likely to follow the reference, so we will get away with this for sure.}. 

\subsection{Citrus}
Extragalactic observers have already catalogued at least one lemon, VV786 \citep{VV77}. Finding citrus fruit in our galaxy is therefore trivial.

\subsection{Tonic}
Gin without tonic is literally unthinkable\footnote{Martinis do exist (Hogg, private communication) but an extensive survey revealed that any bar that can make a decent martini can also produce a G\&T, but many places which can make a G\&T do not produce what the authors consider an acceptable martini. G\&Ts are clearly an earlier, more fundamental stage in the dipsoevolutionary process.}, so we have until this point in the paper simply assumed tonic exists. Tonic without gin is just a shame, and thus the detection of quinine on an exoplanet will inevitably imply the presence of gin. It is a well known fact in exoplanet science that any molecule of interest will be sure to be present and, at least theoretically, detectable in the atmosphere, therefore we estimate the detection thresholds for fluorescing quinine in the atmosphere of exoplanets. 

As any cocktail bartender who favours a cheap trick knows\footnote{Do not be fooled; such gimmicks detract from a good G\&T.}, quinine fluoresces when exposed to UV light; our signature is thus an excess flux in planet spectrum at particular wavelengths. The UV absorption of quinine peaks around 350 nm, and it is important to note that there is no overlap with UV absorption of citric acid components. Average stirring of a typical gin and tonic is not expected to be vigorous enough to produce a double peaked spectrum, which would indicate rapid rotation. The
fluorescent emission peaks at around 460 nm (bright blue/cyan hue), so we should expect a distinctive emission/absorption profile from planets close to the inner edge of the RHZ. 

Noting that the expected absorption is within the Johnson-Cousins U band, and the emission from fluorescence is in the B band, we suggest that the excess B-band flux to absorbed U-band flux ratio be adopted as a critical diagnostic for life. This is defined as 
\begin{equation}
\frac{B_{\rm xs}}{U_{\rm abs}} = \frac{(B_{\rm observed} - B_{\rm expected})}{(U_{\rm expected} - U_{\rm observed}},
\end{equation}
where $B_{\rm expected}$ and $U_{\rm expected}$ are derived from models of the host star flux that no one really understands, but blindly use anyway because stars are hard \citep[see e.g.][ for an overview of all of the ways that we currently inadequately model stars]{Kippenhahn2012}. Any planet with a $B_{\rm xs}/U_{\rm abs}$ ratio close to that predicted by the quantum efficiency of quinine fluorescence (0.58) within the RHZ can be assumed to be inhabited. We encourage all astrobiologists to consider carefully the implications of this B/U Luminosity Life-Sensitive Habitability Indicator Test. 

The presence of this critical signature in the Ultraviolet region of the spectrum highlights the imperative need to urgently invest in a large, UV-capable space telescope such as that proposed in \citet{LUVOIR}. If cost constraints are encountered in the planning and design of such a mission - unlikely, as flagship telescopes are always delivered on time - then we note that for the purposes of finding a drink it is possible to neglect optical and infrared channels. In this case, the ultraviolet only telescope could be named LUSH, the Large Ultraviolet Space Hunter.

\section{Conclusion}

We have considered the likely universal abundance of gins and tonic, deriving for the first time the `Really Habitable Zone' in which life would be worth living. We suggest that efforts should be directed in the near future towards investigating only those planets whose orbits lie within the RHZ, and made unverified claims about the possibility of detecting relevant features. We're off for a drink.

\section*{Acknowledgements}

The authors thank Jake Taylor for proofreading the manuscript with a more sober eye than we had when writing, and Daina Bouquin for finding the Lemon galaxy. We would also like to thank bartenders too numerous to name, especially at Raoul's, for `inspiration', and at Sandy’s wine bar, whose range opens up the possibility for an unlikely follow up paper. 

%% For this sample we use BibTeX plus aasjournals.bst to generate the
%% the bibliography. The sample63.bib file was populated from ADS. To
%% get the citations to show in the compiled file do the following:
%%
%% pdflatex sample63.tex
%% bibtext sample63
%% pdflatex sample63.tex
%% pdflatex sample63.tex

\bibliography{gandt}{}
\bibliographystyle{aasjournal}

%% This command is needed to show the entire author+affiliation list when
%% the collaboration and author truncation commands are used.  It has to
%% go at the end of the manuscript.
%\allauthors

%% Include this line if you are using the \added, \replaced, \deleted
%% commands to see a summary list of all changes at the end of the article.
%\listofchanges

\end{document}